\documentclass{article}

\usepackage{arxiv}
\usepackage{authblk}
\usepackage[utf8]{inputenc}
\usepackage[T1]{fontenc}
\usepackage{hyperref}
\usepackage{url}
\usepackage{booktabs}
\usepackage{amsfonts}
\usepackage{nicefrac}
\usepackage{microtype}
\usepackage{lipsum}
\usepackage{graphicx}
\usepackage{subcaption}
\usepackage{doi}
\usepackage{url}

\title{A commercial embedded board as DAQ in Nuclear Physics}

\author[*]{ M\u{a}d\u{a}lin I. CHERCIU}
\author[*]{ Ciprian M. MITU}
\author[**]{ Ioana ION}
\affil[*]{Institute of Space Science - ISS, Magurele, ILFOV, ROMANIA}
\affil[**]{National Institute for R\&D in Electrical Engineering - ICPE-CA, Bucharest, ROMANIA}

\begin{document}
\maketitle

\begin{abstract}
	The aim of this work is to test the capacity of a low cost but powerful device  to be used in gamma spectrometry. The IZIRUNF0 \cite{izif0-soft} is an embedded STM32 board device with an ARM Cortex-M0 microcontroller build in. The possibilities offered by the configuration (direct memory access controller (DMA) acquisition through circular buffer, 12-bit analog to digital converter (ADC), analog watchdog for threshold on ADC, embedded data acquisition, etc.) make it interesting to be used within an autonomous compact particle spectrometry system. Therefore, we have tested it for a pulse like shape (15 $\mu$s wide) at different frequency, as well as the response from a gamma source Cs137 on a CsI(Tl) scintillator coupled to a photodiode detector. The data were offline processed using our PYTHON \cite{pang} developed program \cite{fit-soft}. 
	  
\end{abstract}


\section{Introduction}
One of the important aspect of the nuclear detector systems is the DAQ, that read and digitize (through analog to digital converters - ADC) the analog signals from the detectors. The faster it is a ADC system the better performance of the system we got (i.e. maximum acquisition rates). DAQ for nuclear detectors can be dependent by others computing system (MC DT3034 \cite{mc1}, ComTec GmbH MCA 3FADC \cite{fastcomtec2}) or performed autonomous [MC USB-1602HS/1604HS \cite{mc2}, ORTEC EASY-MCA \cite{ortec1}, ComTec GmbH MCA 8000A \cite{fastcomtec1}], but quite expensive.
The most important aspects of the development board IZIRUNF0 under test are: rapid ADC conversion time(through DMA), small dimensions, on board data acquisition and low power consumption that make it an interesting tool to be used in various applications (portable gamma monitoring, autonomous monitoring, etc.). 

The IZIRUNF0 utilized an extended version of Little Kernel \cite{lk-soft}, developed by IZITRON \cite{izitron-web} for their boards IZIRUNF 0\( \sim \)7. We have used a specific branch of izirunf0 software \cite{izif0-soft} that create threads for each task (system load, daq, etc.). The processing, filtering, signal recognition are done offline to maximize the acquisition rate. Furthermore, because the IZIRUNF0 board has a flash memory of only 512 kB, we have used an external FLASH memory of 16 GB to be used for large data acquisition. The data can be access directly from SD card or send it through SPI port (bluetooth or wifi). 

IZIRUNF0 is part from IZIRUN boards serie, based on STM32(F0, F4, F7). All the boards use M.2 67-pins allowing external communication through I2C, SPI, CAN and UART using IZIGOBOARD SHIELD. The boards are powered by ARM Cortex(-M0, -M4, -M7) and include built-in peripherals like flash, eeprom, quatz, leds, and buttons. Nevertheless, using IZIGOBOARD SHIELD it is also possible to access the peripherals like Ethernet and GPIO, and additional functionality like USB communication. 

\begin{figure}[!ht]
	\centering
	\begin{subfigure}[b]{0.27\linewidth} 
		\includegraphics[width=\linewidth]{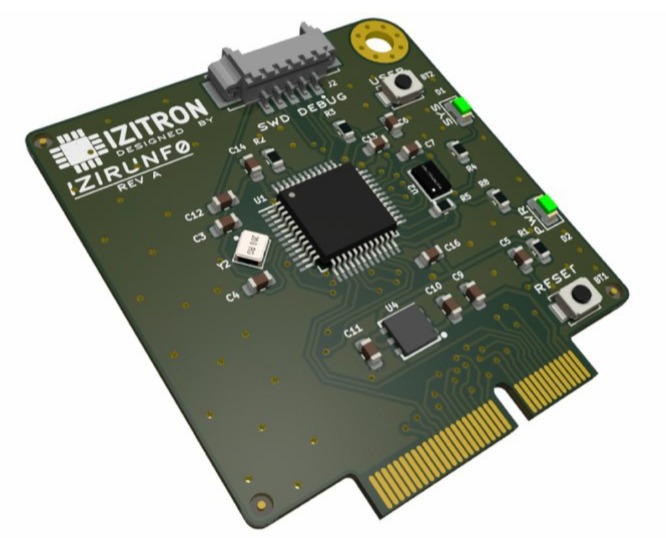}
		\caption{perspecive view}
	\end{subfigure}
	\begin{subfigure}[b]{0.45\linewidth}
		\includegraphics[width=\linewidth]{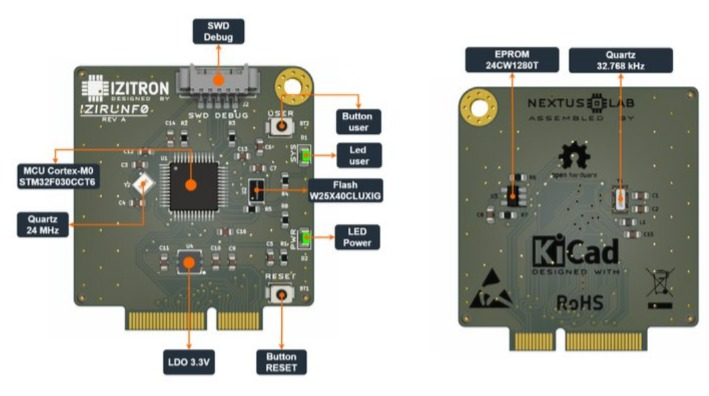}
		\caption{top/bottom view}
	\end{subfigure}
	\caption{IZIRUNF0 [\ref{fig:izirunf0}] development board}
	\label{fig:izirunf0}
\end{figure}

Our work is based on IZIRUNF0 [\ref{fig:izirunf0}] that has ARM Cortex-M0 at 48MHz max as MCU, 512 kB FLASH, 32 kB RAM, 2-UART, 1-I2C, 1-SPI, 4-ADC, 6-PWM, 23-GPIO and is designed for standalone low power applications and IoT projects.

\section{DAQ of the system}
\label{sec:daq}
\begin{figure}[!ht]
	\centering
	\begin{subfigure}[b]{0.43\linewidth} 
		\includegraphics[width=\linewidth]{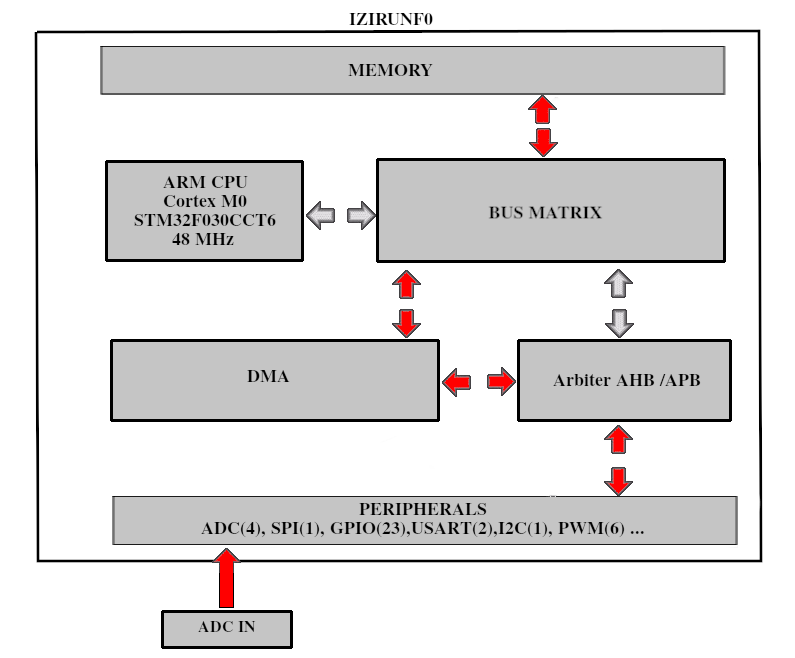}
		\caption{DAQ - first level}
		\label{fig:daq1}
	\end{subfigure}
	\begin{subfigure}[b]{0.4\linewidth}
		\includegraphics[width=\linewidth]{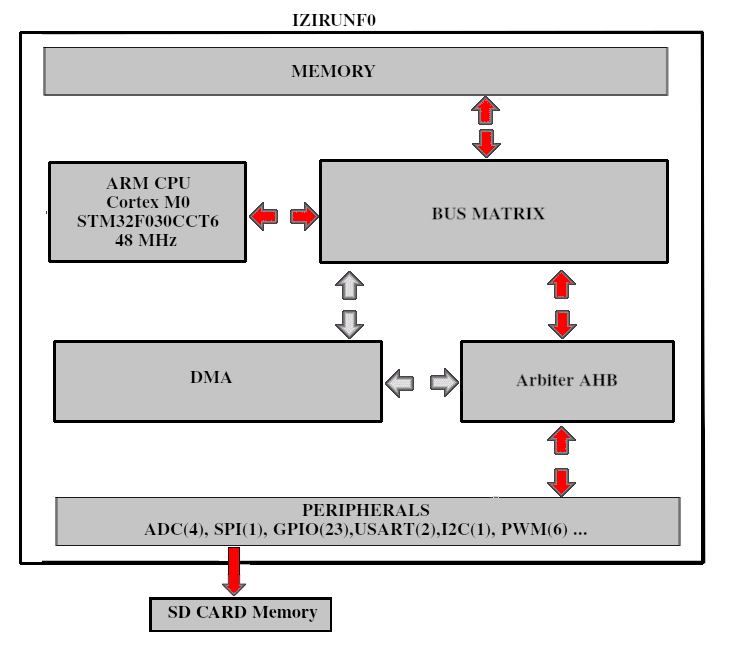}
		\caption{DAQ - second level}
		\label{fig:daq2}
	\end{subfigure}
	\caption{DAQ diagram from the analog signal received to collected on a SD card}
	\label{fig:daq}
\end{figure}

The data acquisition diagram is presented in the Fig.~\ref{fig:daq} where it is show the way the signal is collected from the analog input and written to memory. We choose to write the data to the SD card memory because of the increased speed to write to memory, of large size capacity and because it offers the possibility to be used as an autonomous system to acquire the data. 

The analog signal value is continuously checked by the watchdog feature of the converted voltage. The 12-bit analog to digital converter offer the possibility to performs conversions in single-shot or scan modes. The second option means an automatic conversion is performed on the analog input. 

The ADC converted values are continuously written, through DMA, to the memory buffer (see figure \ref{fig:daq1}). The buffer size has to be chosen accordingly to the pulse width. In our case we have choose 60 ADC sample values per buffer. The speed and accuracy of values are directly related to the sampling time ($t_{S}$) or cycles which varies from 0.11 to 17.1 $\mu s$ and so a total conversion time ($t_{CONV}$) we got is between 0.3 to 5.2  $\mu s$ (8 bits to 12 bits precision). 

\begin{figure}[!ht]
	\centering
	\begin{subfigure}[b]{0.45\linewidth}
		\includegraphics[width=\linewidth]{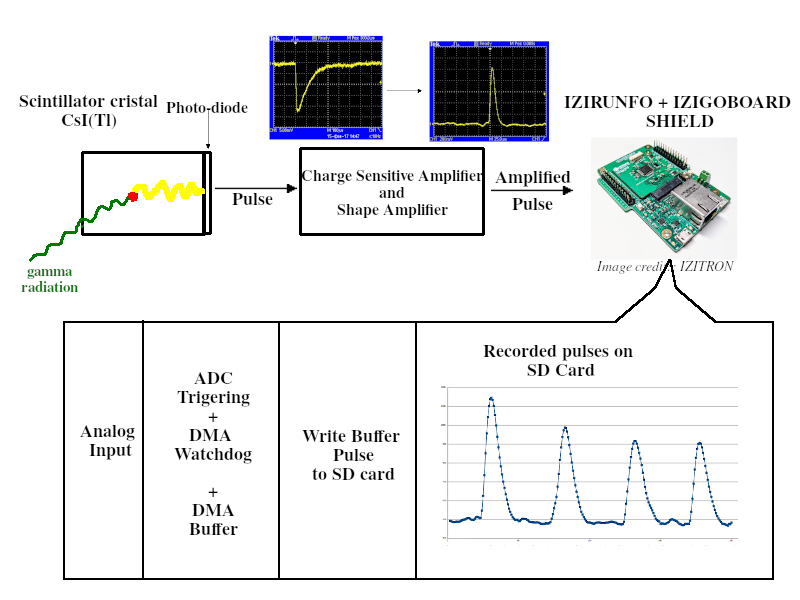}
		\caption{DAQ system for analog amplified detector pulse \newline (scintillator cristal and photo-diode)}
		\label{fig:daq_source}
	\end{subfigure}
	\begin{subfigure}[b]{0.45\linewidth}
		\includegraphics[width=\linewidth]{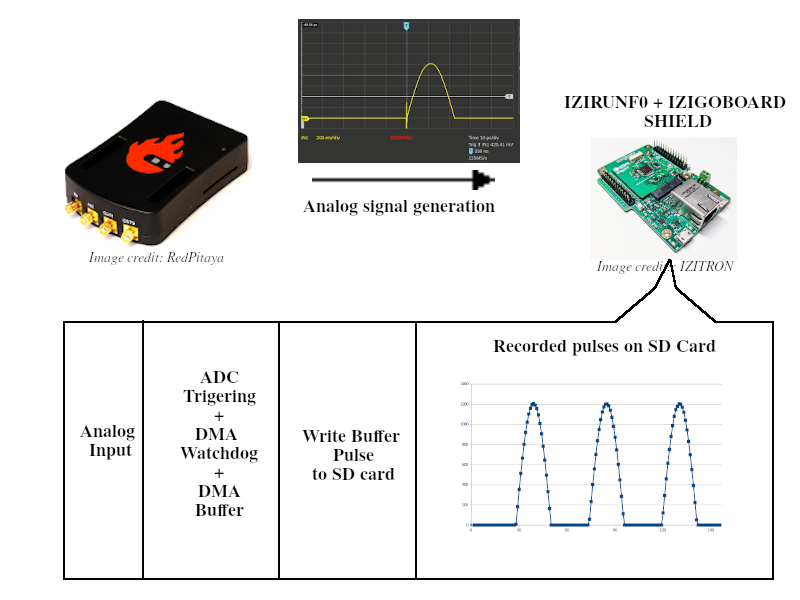}
		\caption{DAQ system for analog signal generation using
		 RedPitaya}
		\label{fig:daq_redpitaya}
	\end{subfigure}
	\caption{DAQ system with IZIRUNF0}
	\label{fig:process1}
\end{figure}

More cycles on a conversion mean higher ADC conversion precision but this has to be chosen accordingly with the pulse characteristics. In our case we have used the faster conversion because our pulse is sharp (15 - 25) $\mu$s. 
When a pulse arrives to the analog input, the signal is converted in digital as fast as possible (accordingly to the ADC conversion cycles) the watchdog continuously check if the ADC converted value exceed the threshold value. When the threshold condition occurs, a signal interrupt is called. This is the moment when a flag is raised that a high pulse is on. Using a developed version of the Little Kenel(LK) \cite{izif0-soft}, the buffer data is written to SD card formated as FAT32 (see figure \ref{fig:daq_redpitaya}).
The ADC converted values, from the buffer, are written to the memory SD card (see figure\ref{fig:daq2}), as binary file. The dimension of the file depend on the buffer size, acquisition time, incoming pulse frequency, threshold level, etc. 

In the figure \ref{fig:process1} is shown the steps followed after a pulse arrives at the analog input pin on board card. The board has been tested using RedPitaya \cite{redpitaya} (see figure \ref{fig:daq_redpitaya}) and using a scintillator + photo-diode detector \ref{fig:daq_source} .

\section{Data processing}
\label{sec:processing}

\begin{figure}[!ht]
	\centering
	\begin{subfigure}[b]{0.49\linewidth}
		\includegraphics[width=\linewidth]{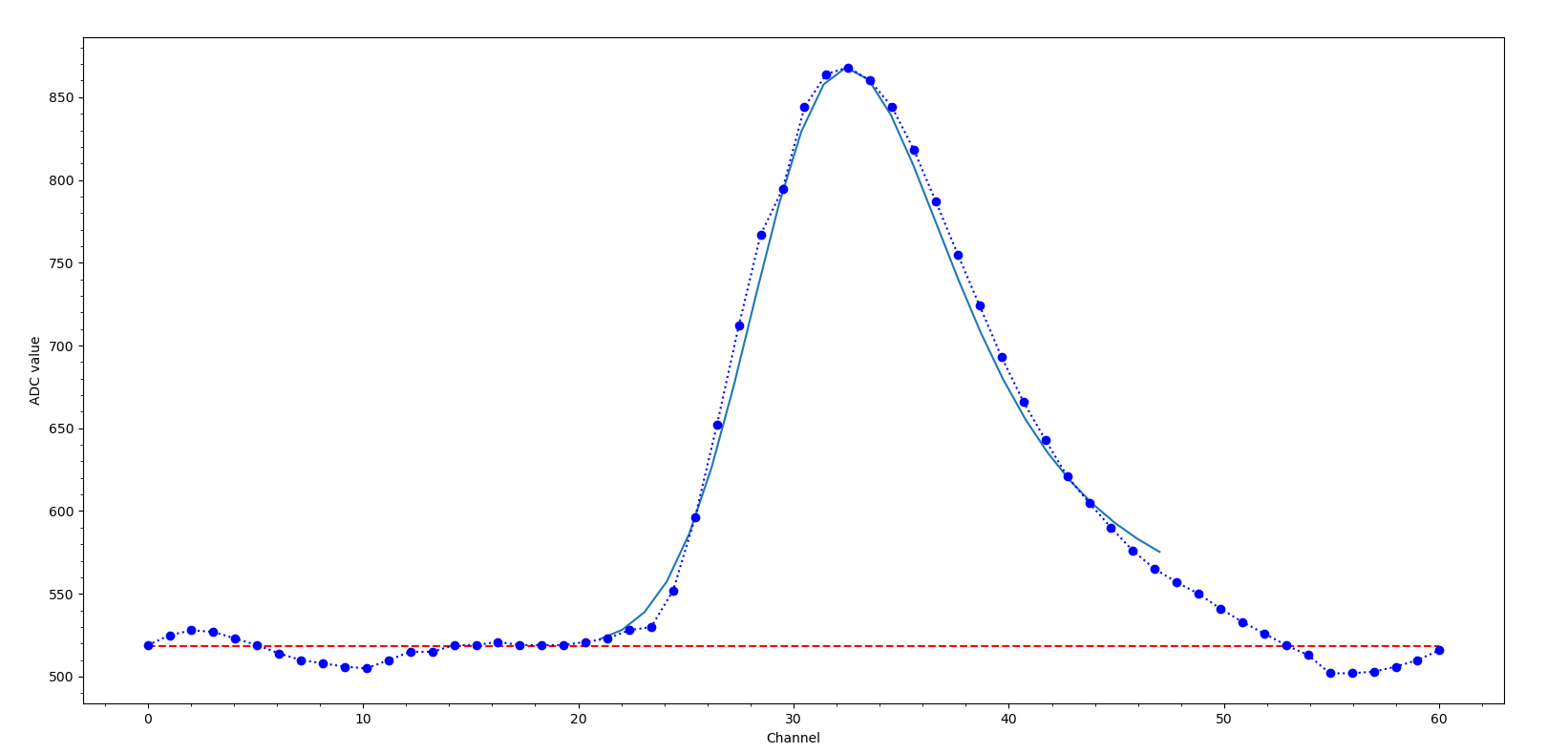}
		\caption{One pulse recognition}
		\label{fig:pulse}
	\end{subfigure}
	\begin{subfigure}[b]{0.49\linewidth}
		\includegraphics[width=\linewidth]{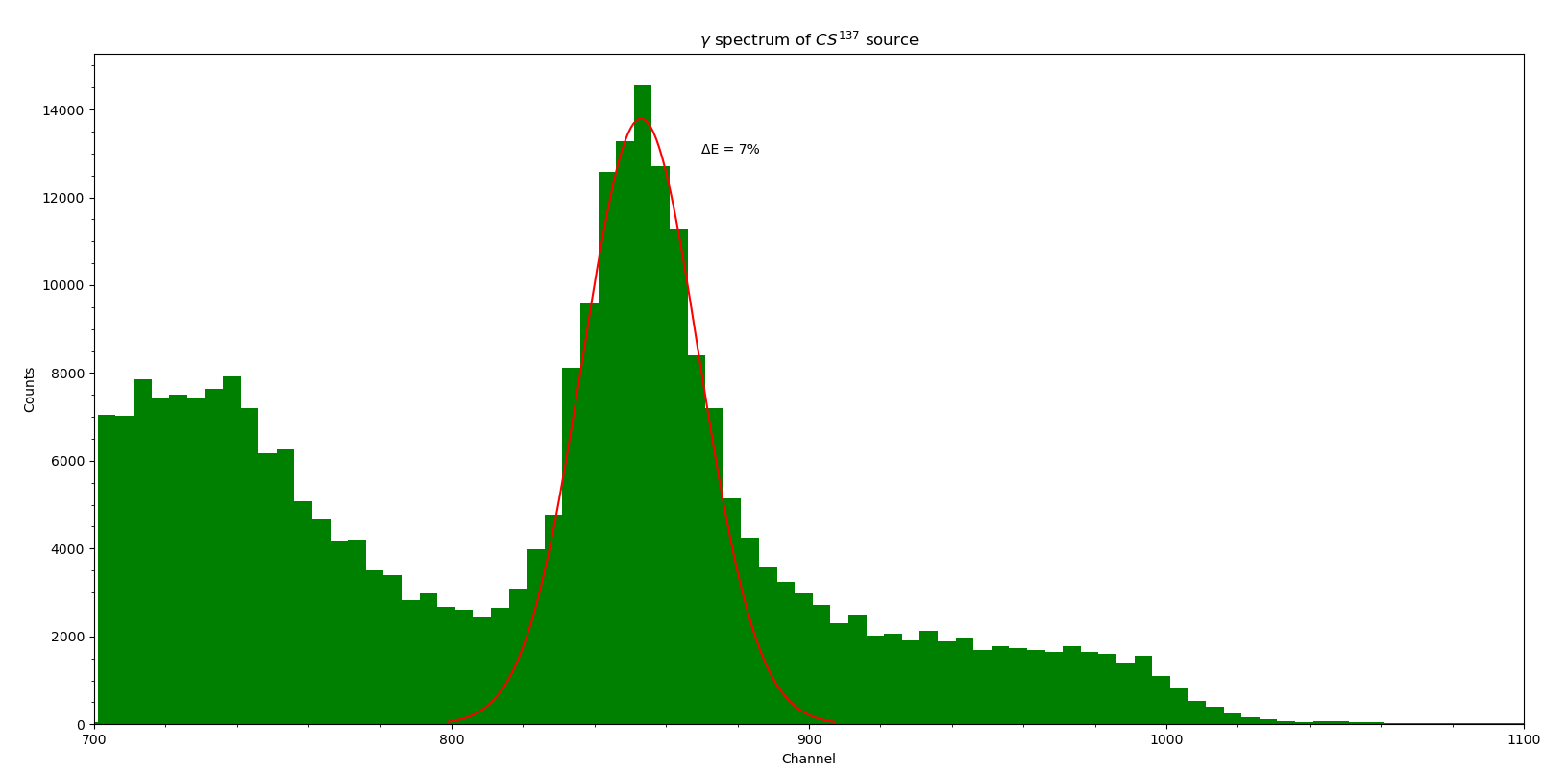}
		\caption{$\gamma$ energy spectrum of $Cs^{137}$}
		\label{fig:spectrum}
	\end{subfigure}
	\caption{Offline analysis }
	\label{fig:process2}
\end{figure}

We have tested our offline processing software using a Cs137 gamma source. After the acquisition complete, we had read the data from SD card. The IZIRUNF0+IZIGOBOARD board, holds two UART connections and, if a wireless connection is recquired, we can use the second UART connection to access the acquired data because the first one is used by the command line communication purpose.
In the offline analysis, the pulses were recognized, fitted (see figure \ref{fig:pulse}) and performe pulse high analysis (PHA) for energy spectrum distribution (see figure \ref{fig:spectrum}) using our developed PYTHON program \cite{fit-soft}. A resolution better then 7$\%$ is obtained for characteristic gamma energy peak at 661.62 keV of $Cs^{137}$ radionuclide. 

\section{Results and conclusions}
\label{sec:results}

\begin{figure}[!ht]
	\centering
	\begin{subfigure}[b]{0.80\linewidth}
	\includegraphics[width=\linewidth]{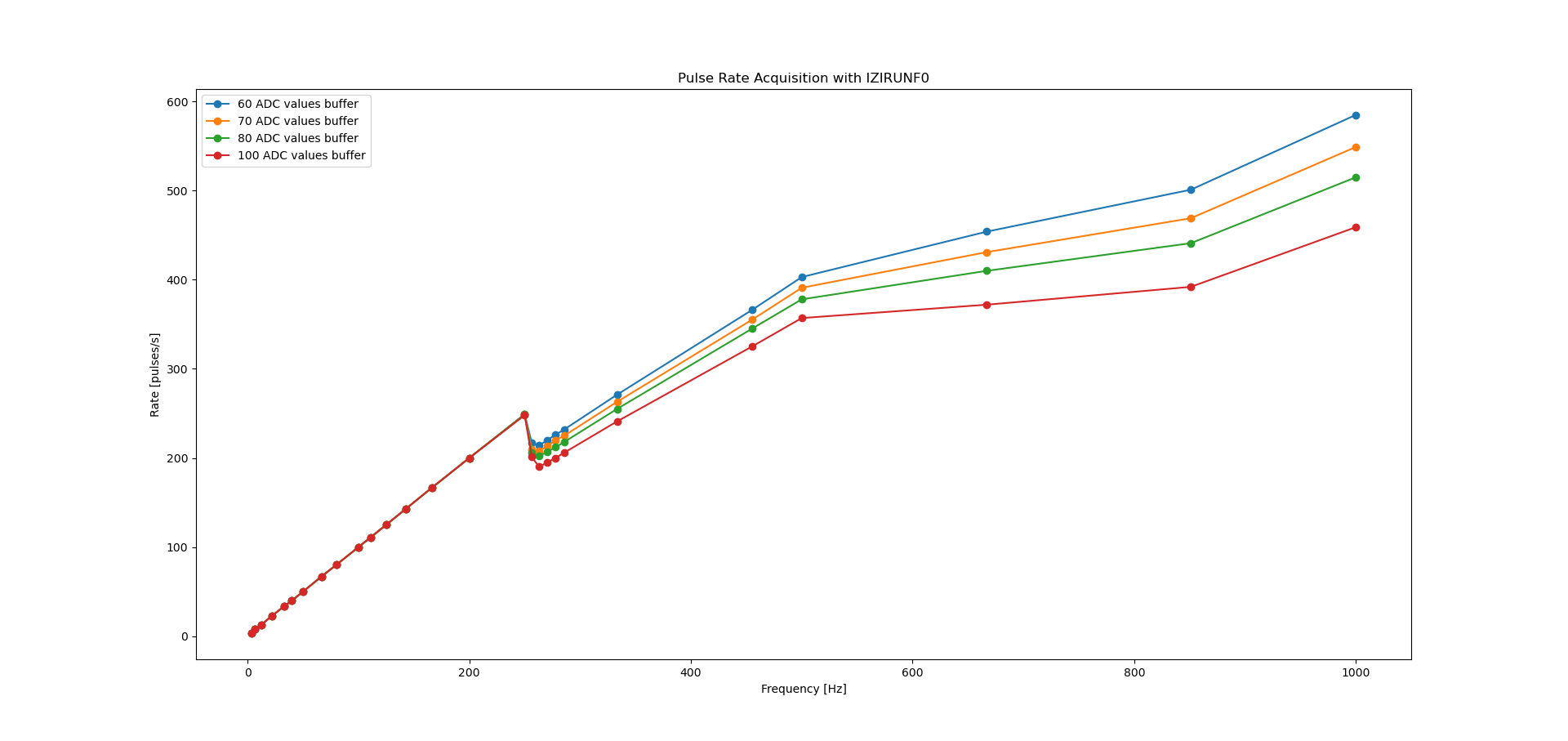}
\end{subfigure}
\caption{Pulse rate acquisition by MCU-board izirunf0 \\
as generated signal with RedPitaya \cite{redpitaya}. }
\label{fig:pulse_rate}
\end{figure}

\begin{figure}[!ht]
	\centering
	\begin{subfigure}[b]{0.80\linewidth}
		\includegraphics[width=\linewidth]{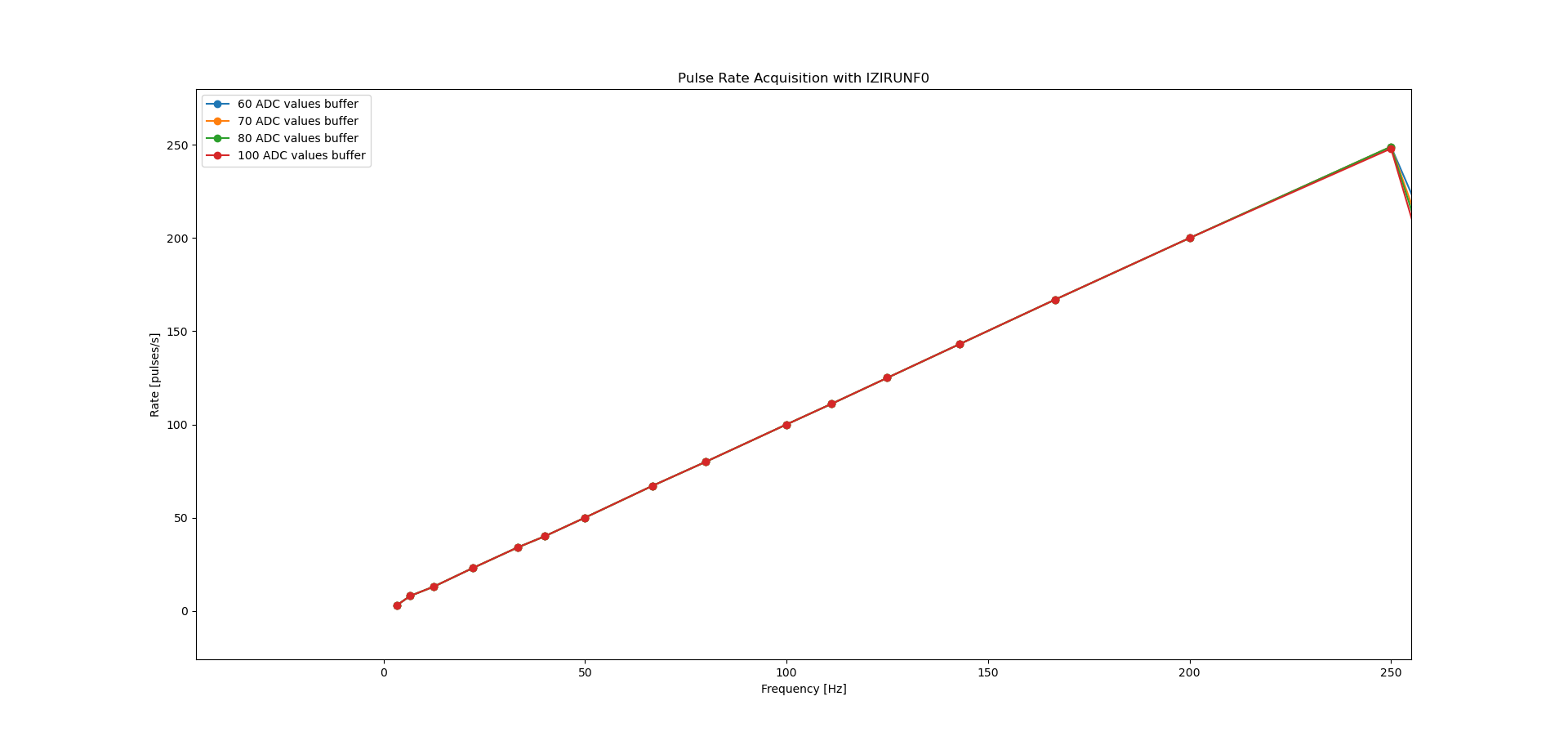}
	\end{subfigure}
\caption{Zoom on linear segment of the pulse rate acquisition by MCU-board. }
\label{fig:rate_zoom}
\end{figure}

We have tested the MCU based board as rapid data acquisition system for gamma spectroscopy. Using STM32 based board on can access directly, through DMA, peripheral to memory, memory to peripheral or memory to memory information and we have tested how fast and reliable is such a device for gamma spectroscopy data acquisition. 
In the figure \ref{fig:pulse_rate} we can see the response of the IZIRUNF0 board as a function of a generated signal frequency, a gaussian like shape  (see in figure \ref{fig:daq_redpitaya}) using RedPitaya signal generator. 
The linear segment (see \ref{fig:rate_zoom}), from 3 to 250 pulses/s, has an identical rate acquisition as generated signal and represent the linear response of the board. Acquiring data on an external SD card we obtain a maximum 250 pulse/s (15 kHz sampling rate) in a reliable manner. The board can acquire even faster (as much as 1 MHz) the data but, a bottleneck done by 512 bits sector size of FAT32 file system, limits the acquisition data within this range. Within such configuration, acquiring data at higher frequency result in loosing some pulses. That happen because the write function has to close the sector and start another one and the period of the pulse is to short than the time required for the write function to return.

Furthermore, we also verified if the buffer dimension plays a role on the acquisition data rate. As we already mentioned, the buffer dimension is a matter of shape and width of the pulse and doesn't affect the acquisition rate within the reliable region. Even so, we can see a worse data acquisition as the buffer increase, from 60 ADC converted values up to 100 ADC values, for higher frequency larger then 250 pulses/s (see figure \ref{fig:pulse_rate}).

\begin{figure}[!ht]
	\centering
	\begin{subfigure}[b]{0.80\linewidth}
		\includegraphics[width=\linewidth]{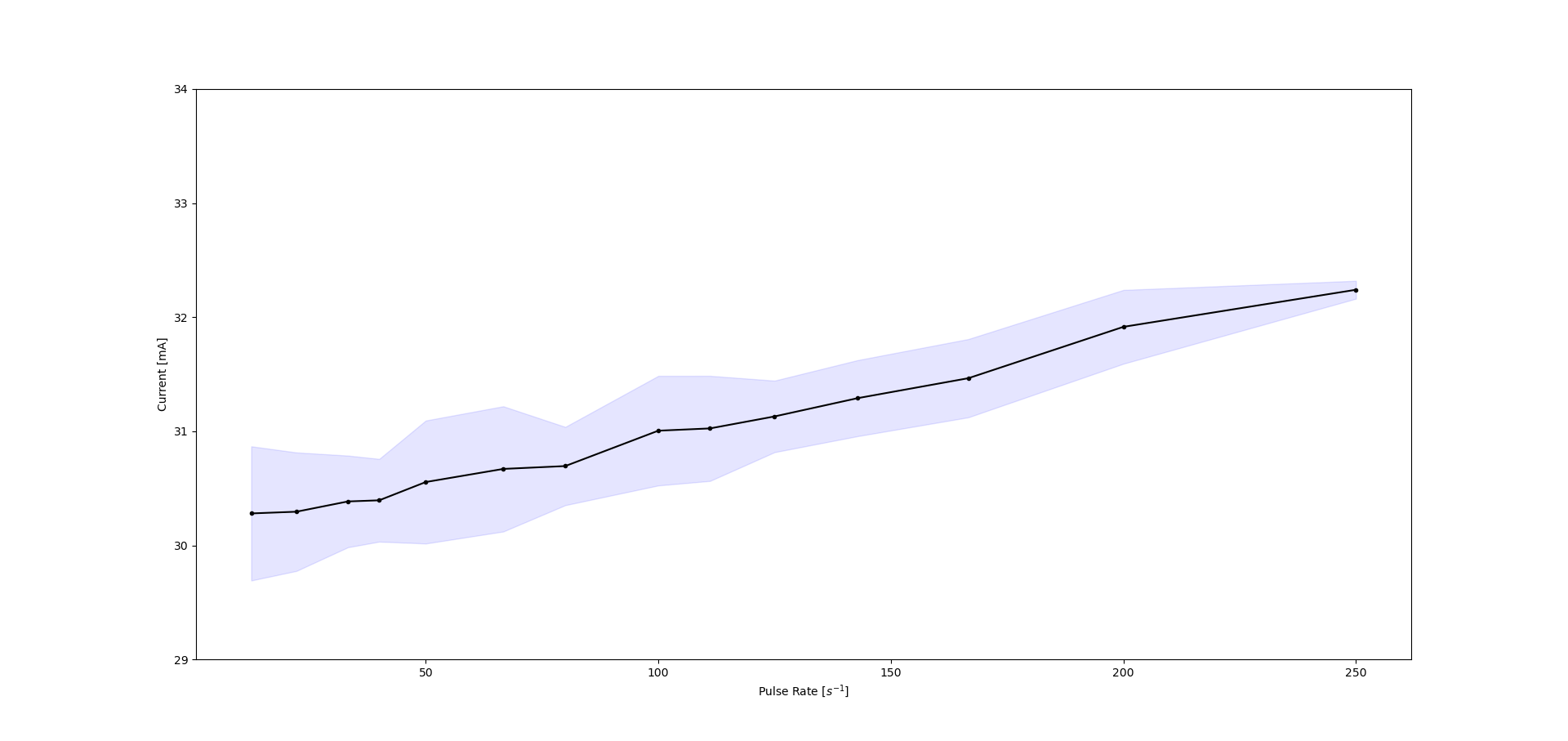}
	\end{subfigure}
\caption{Current consumption as a function of pulse rate acquisition on SD}
\label{fig:rate_mA}
\end{figure}

Another aspect that we looked for was the current consumption on different acqusition rate. We found that the current used by the configuration (IZIRUNF0 board + external SDcard ) is within the range 30 - 32 mA \ref{fig:rate_mA} which is quite resonable for such, not specific, embedded board data acquisition. However, the board can be drawn in sleep mode were the power consumption drops significantly, but this is not the case when continuously acquisition is in progress.     

The IZIRUNF0 board is an interesting board even for such a task as rapid data acquission for gamma spectrometry and we have obtained with only a 3 $cm^3$ cristall scintillator a resolution o 7$\%$.
It acquire correctelly up to 250 pulses/s (15-25 $\mu$s wide) on an external SD card but increasing the internal flash from 512 kB to, at least, few hundred of MB, will increase the acquisition rate at least one order of magnitude, however the maximum ADC sampling rate is quite large few MHz. An upgrade of the board, especially an upgrade of the flash, a wifi/bluetooth build in connection will make it a precious board for high rate data acquisition. The board is light, small size, low power consumption, 3.3/5V DC input tolerant, large number of peripheral connection that make us considerate it an interesting DAQ board for autonomous data aquisition.         

\section{Acknowledgements}
\label{sec:acknowledgements}
We would like to thanks to Ph.D Elena Stancu from STARDOOR Laboratory of National Institute for Laser, Plasma and Radiation Physics (INFLPR) - Romania, to give us the possibility to acquire energy spectrum of $\gamma$ radiation, using low activity commercial source of the $Cs^{137}$.

\newpage
\bibliographystyle{prsty}
\bibliography{references}

\end{document}